\documentclass[preprint,12pt]{elsarticle}
\usepackage{graphicx}
\usepackage{amssymb}
\usepackage{amsthm}
\usepackage{lscape}
\usepackage{mathtools}
\usepackage{color}
\usepackage{makecell}

\journal{Radiation Physics and Chemistry}

\begin{document}
\begin{frontmatter}


\title{Evaluation of accurate uncertainty of measurement in L subshell ionization cross-section}

\author[label1]{Shashank Singh}
\author[label2]{Soumya Chatterjee}
\author[label2]{D. Mitra}
\author[label3]{T. Nandi}


\address[label1]{Department of Physics, Panjab University, Chandigarh-160014, India.}
\address[label2]{Department of Physics, University of Kalyani, Kalyani, West Bengal-741235, India.}
\address[label3]{1003 Regal, Mapsko Royal Ville, Sector 82, Gurugram-122004, Haryana, India.
correspondence:\hspace{0.1cm} nanditapan@gmail.com\\Superannuated from Inter-University Accelerator Centre, Aruna Asaf Ali Marg, New Delhi-110067, India.}




\begin{abstract}
To have a better understanding of a physical process, a comparison of experimental data with theoretical values is mandatory. The comparison is meaningful if the uncertainty in the experiment is accounted well. However, it is seldom seen, especially for a complex phenomenon. We take a test case through L subshell ionization of atoms by particle impact. \textcolor{red}{Experimentally, x-ray production cross-sections are measured, but ionization cross-sections are calculated theoretically.} Furthermore, the uncertainty of the x-ray production cross-section is mainly statistics and detector-efficiency driven. But ionization cross-section involves many other factors because of the relationship between the production and ionization cross section, having wide uncertainty spectrum. Consequently, determining the measurement uncertainty in L subshell ionization cross-section is always difficult. We have studied this issue in the simplest way, where the rule of weighted propagation of relative uncertainty is utilised. We notice that larger uncertainties are involved in atomic parameters relevant to L$_1$ (2s$_{1/2}$) subshell than those associated with the other two L$_2$ (2p$_{1/2}$) and L$_3$ (2p$_{3/2}$) subshells. \textcolor{red}{Hence, comparison between theory and experiment would give higher emphasis on $L_2$ and $L_3$ subshell ionization cross sections.} We believe this work aware us that the appropriate uncertainty evaluation is extremely important for providing the right judgment on the data.
\end{abstract} 


\end{frontmatter}


\section{Introduction}
\label{S:1}
\indent According to the “International Vocabulary of Basic and General
Terms in Metrology" a measurement is defined as a “set of operations having the
object of determining a value of a quantity” \cite{dybkaer2011definitions}. Therefore, a result of a measurement is an estimate of the value of a measurand and should be accompanied by an uncertainty statement \cite{choi2003uncertainty}. The uncertainty of measurement is the doubt that exists about the result of any measurement. But for every measurement - even the most careful - there is always a margin of
doubt. It is a complicated subject, and still evolving. So there is a great need for a guide that provides clear, down-to-earth explanations, easy enough for non-expert readers \cite{bell2001beginner}. \textcolor{red}{While error is the difference between the measured value and the ‘true value’ of the quantity being measured.} The ‘true’ value can only be found from a very large set of measurements and thus any uncertainty whose value is not known is a measure of uncertainty. Hence, only an estimate of the standard deviation can be found from a moderate number of values and one standard deviation is an estimate of the uncertainty. The “Guide to the Expression of Uncertainty of Measurement” (GUM) provides the general rules for quantifying the uncertainty in a measurement \cite{bipm2008evaluation}. As a rule of thumb usually between 4 and 10 readings is sufficient to make a good quality measurement and its uncertainty. \\
\indent The uncertainty evaluation is an important part of a good quality measurement. It is only
complete if it is accompanied by a statement of the uncertainty. A measurand $Y$ is determined
from N other quantities $X_i$ using a functional relationship as follows
\begin{align}
Y = f(X_1, X_2, …, X_N)  
\end{align}
where the quantities $X_1, X_2, …, X_N$ may themselves be viewed as measurands and may depend on other quantities like $X_i=f(x_i)$. Two approaches based on the evaluation method are used to estimate the standard
uncertainty: Type A evaluation is based on statistical treatment on a series of observations, and Type B evaluation is based on non-statistical one such as data from manufacturer’s specifications, previous experimental data, theoretical data, etc. Note that both Type A and Type B uncertainties can be due to a “random effect” as well as to a “systematic effect” in nature. Random effect is recognizable from the variations in the repeated observations, where uncertainty analysis of the observed data is governed by the rules of statistics. On the other hand, though the contribution of systematic effect appears in the repeated measurements, but we can learn nothing from these. Other methods are needed to estimate the uncertainties due to systematic effects.\\
\indent Inner shell ionization of an atom is a very interesting field of study for the theoreticians as well as experimentalists since last century. For a better understanding of the inner shell ionization process, a lot of work had been reported using different experimental pathways. In this endeavour, inner shell ionization induced by electron \cite{burhop1940inner,rester1966k,PhysRevA.11.1475,PhysRevLett.39.82,PhysRevA.77.042701,LIU20191,zhao2020measuring,tian2020measurements}, photon \cite{storm1967photon,VEIGELE197351,PhysRevA.31.2918,doi:10.1002/xrs.1300180503,doi:10.1002/xrs.1300230505,ertuugrul1996measurement}, proton \cite{PhysRev.165.66,garcia1970inner,PhysRevA.29.70,PhysRevA.35.2007,CHEN1989257,PhysRevA.42.261,MONTANARI2017236,MIRANDA2019,PEREZ201921} and varieties of heavy ions \cite{PhysRevLett.29.1361,ANDERSEN1976507,PhysRevA.30.2234,SARKAR199523,PhysRevA.30.2234,singh2000subshell,PhysRevA.73.032712,MSIMANGA201690,VERMA201830,GORLACHEV201831,chang2018multiple,oswal2018x,SILHADI2019158,OSWAL2020108809} had been reported. To explain the ionization processes, various theories like SCA \cite{KOCBACH1980281}, PWBA \cite{merzbacher1958handbuch}, ECPSSR \cite{brandt1979shell, brandt1981energy,lapicki2002status, PhysRevA.20.465,PhysRevA.23.1717}, ECUSAR \cite{lapicki2002status} and SLPA \cite{PhysRevA.67.062901} have been used. In a lot of studies, comparisons were made between the experimental cross-section data and corresponding theoretical predictions. Accuracy of experimental data is very important. Given uncertainty alongside the data gives the accuracy of such measurement. Finite uncertainty in the x-ray production cross-section data is indispensable due to the uncertainty in photo-peak area evaluation, efficiency of the detector, ion beam current, target thickness, etc. The ionization cross-sections can be extracted from the measured x-ray production cross-sections using suitable atomic parameters. Uncertainties in the x-ray production cross-sections and the used atomic parameters are propagated into the ionization cross-sections.  Conversion of K shell x-ray production cross-section into the ionization cross-section requires only one atomic parameter, fluorescence yield \cite{merlet2004measurements}, hence evaluating the uncertainty is simple. In contrast, obtaining the L subshell ionization cross-section from the x-ray production cross-section data requires several atomic parameters including fluorescence yield, Coster-Kronig yield, the fraction of radiative transition rates, which in turn introduce the Type B uncertainty.  Thus, the calculation of uncertainty is complex and significant discrepancies prevalent in literature. To examine the reason behind such large variation, we have made a thorough study of the propagation of uncertainties from the L x-ray production cross-sections to the L subshell ionization cross-section. Details will be seen in the next section, where the method of weighted uncertainty propagation rule has been applied. For example, we have taken results from recent work \citet{OSWAL2020108809} and assessed the quoted accuracy. More so, an attempt is put on suggesting the best possible way to perform the most accurate experiment. Probably, besides the L subshell ionization, this work will be a general guideline for uncertainty analysis of many other measurements relevant to physics or beyond.\\

\section{Present scenario of measurement uncertainty in L-shell ionization studies}
In the past, many experiments have studied the L-shell ionization in various targets bombarded by different projectiles in certain energy ranges and quoted the measurement uncertainties. A brief list is given in the table \ref{S:2}. Though the experiments have followed similar detection systems, but the measurement uncertainties vary in a wide range of 4-35\%. This scenario is a deterrent for improving the theoretical understanding. 
\label{S:2}
\begin{landscape}
\begin{table}
\tiny
\caption{\label{tab:table1} Uncertainties quoted for L shell production and ionization cross sections in earlier experiments.}
\centering
\begin{tabular}{l l l l l l l l l l}
\hline
\textbf{Ref.} & \makecell{\textbf{Targets}}&  \makecell{\textbf{Projectile, its energy}} & \makecell{\textbf{$\delta L_1$}, \textbf{$\delta L_2$}, \textbf{$\delta L_3$ (\%)}} & \makecell{\textbf{Remarks}}\\
\hline
\citet{PhysRevA.19.1930}& \makecell{W} & \makecell{electron, 11-40 keV} & \makecell{17, 15, 15} & \makecell{Nearly equal uncertainties\\  for all subshells}\\
\citet{palinkas1980subshell}& \makecell{Au, Pb, Bi} & \makecell{electron, 60-600 keV} & \makecell{22, 15, 11} & \makecell{7-10\% uncertainties for x-ray production \\ cross sections for all the three elements \\ }\\ 
\citet{reusch1986method}& \makecell{$29\leq Z \leq79$} & \makecell{electron, 50-200 keV} & \makecell{7-15, 6-15, 6-15}& \makecell{Nearly same uncertainties for all the subshells\\ and about 15\% for total L ionization cross section}\\
\citet{PhysRevA.37.106}& \makecell{W} & \makecell{electron, 12-40 keV}&\makecell{--} &\makecell{10\% uncertainty for $\sigma_3/\sigma_2$  and $\sigma_2/\sigma_1$}\\
\citet{llovet2014cross}& \makecell{$1\leq Z \leq99$}&\makecell{electrons and positron, \\up to 1 GeV}& \makecell{10-30, 10-30, 10-30}& \makecell{Similar uncertainties for all elements \\and all subshells}\\
\citet{barros2015ionization}& \makecell{Au}&\makecell{electron, 50-100 keV} &\makecell{22, 11, 12} & \makecell{Without inclusion of uncertainties \\ in atomic relaxation parameters}\\
\citet{barros2015ionization}& \makecell{Au} & \makecell{electron, 50-100 keV}& \makecell{23, 12, 13}  &\makecell{With inclusion of uncertainty in atomic \\relaxation parameters \cite{rao1972atomic,kolbe2012subshell,scofield1974exchange,campbell1989interpolated}}\\
\citet{barros2015ionization}&\makecell{Au} &\makecell{electron, 50-100 keV} & \makecell{28, 16, 16}  &\makecell{With inclusion of uncertainty in atomic \\ relaxation parameters as given in \cite{krause1979atomic,scofield1974relativistic,rao1972atomic}}\\
\citet{bernstein1954shell}& \makecell{Ta, Au, Pb, U} & \makecell{H, 1.5-4.25 MeV} &\makecell{--}& \makecell{20\% Uncertainty for all \\ x-ray production cross-sections}\\
\citet{PhysRevA.11.607}& \makecell{Ta, Au, Bi} & \makecell{H, 1 to 5.5 and \\ He, 1 to 11 MeV} & \makecell{22, 28, 21}& \makecell{Uncertainties for $L_\alpha$ is about \\ 15\%, for $L_{\gamma_{1+5}}$ and $L_{\gamma_{2+3+4}}$ are about 15-18\%}\\
\citet{PhysRevA.15.943}& \makecell{Au, Tl, Pb, Bi, Th, U}& \makecell{H, 0.5-3.5 MeV} &\makecell{--}& \makecell{7\% uncertainty for production cross section\\ and  error bar for ionization cross sections \\given in figures only}\\
\citet{wheeler1979subshell}& \makecell{Pr, Sm, Tb, Ho, Yb} &\makecell{H, 150-400 keV}& \makecell{20, 20, 15} & \makecell{Same uncertainty for all the elements}\\
\citet{sokhi1984experimental}& \makecell{Different targets}&\makecell{H} & \makecell{6-20, 7-20, 6-26}& \makecell{Uncertainty differs with target}\\
\citet{miranda2014experimental}& \makecell{Ne to Am} & \makecell{H, 0 keV to 1 GeV} & \makecell{--}& \makecell{No regular trend seen in uncertainties}\\
\citet{semaniak1994a} & \makecell{Between La to Au}&\makecell{ $^{14}N$, 1.75-22.4 MeV}& \makecell{25-10, 25-15, 25-10} & \makecell{Uncertainties at low energy were \\10-25\%  and 10\% at high energies}\\
\citet{Semaniak1995b}& \makecell{$72\leq Z \leq 90$}&\makecell{C and N, 0.4 to 1.8 MeV/amu} &\makecell{6-20, 5-12, 4-10}& \makecell{Uncertainties differs with target}\\
\citet{dhal1995subshell}& \makecell{Au, Bi}&\makecell{B, 4.8-8.8 MeV} & \makecell{10, 10, 10}& \makecell{Same for both the elements}\\
\citet{banas2002role}& \makecell{Au}&\makecell{O of 0.4-2.2 MeV/amu} &\makecell{15-30, 15-30, 15-30}&  \makecell{10-20\% for all production cross-section}\\
\citet{pajek2003multiple}& \makecell{Au, Bi, Th, U}&\makecell{O, 6.4 to 70 MeV} & \makecell{--}& \makecell{Uncertainties for production cross-sections \\ of $L\alpha_{1,2}$ were 9\%, for $L_{\gamma_1}$ 13\% and \\ for $L_{\gamma_{2,3}}$ 10-25\% and\\  for subshell ionization cross section the \\error bars included in figure only}\\
\citet{fijal2008subshell}& \makecell{Au and Bi}&\makecell{S, 12.8-120 MeV} & \makecell{10-30, 10-30, 10-30}& \makecell{Uncertainties for production cross-sections \\ of $L\alpha_{1,2}$ were 7-15\%, for $L_{\gamma_1}$ 10-18\% and \\ for $L_{\gamma_{2,3}}$ 15-25\% for all elements}\\
\citet{OSWAL2020108809}& \makecell{Ta, Pt, Th, U}& \makecell{Si of 84-140 MeV} & \makecell{15-20 for Ta and 30-35 \\for other targets, 12-15, 12-15} & \makecell{10-12\% for the production cross sections,\\ uncertainties were evaluated \\ quite satisfactorily} \\
\hline
\end{tabular}
\end{table}
\end{landscape}
\begin{figure*}
\includegraphics[width=140mm,scale=01.0]{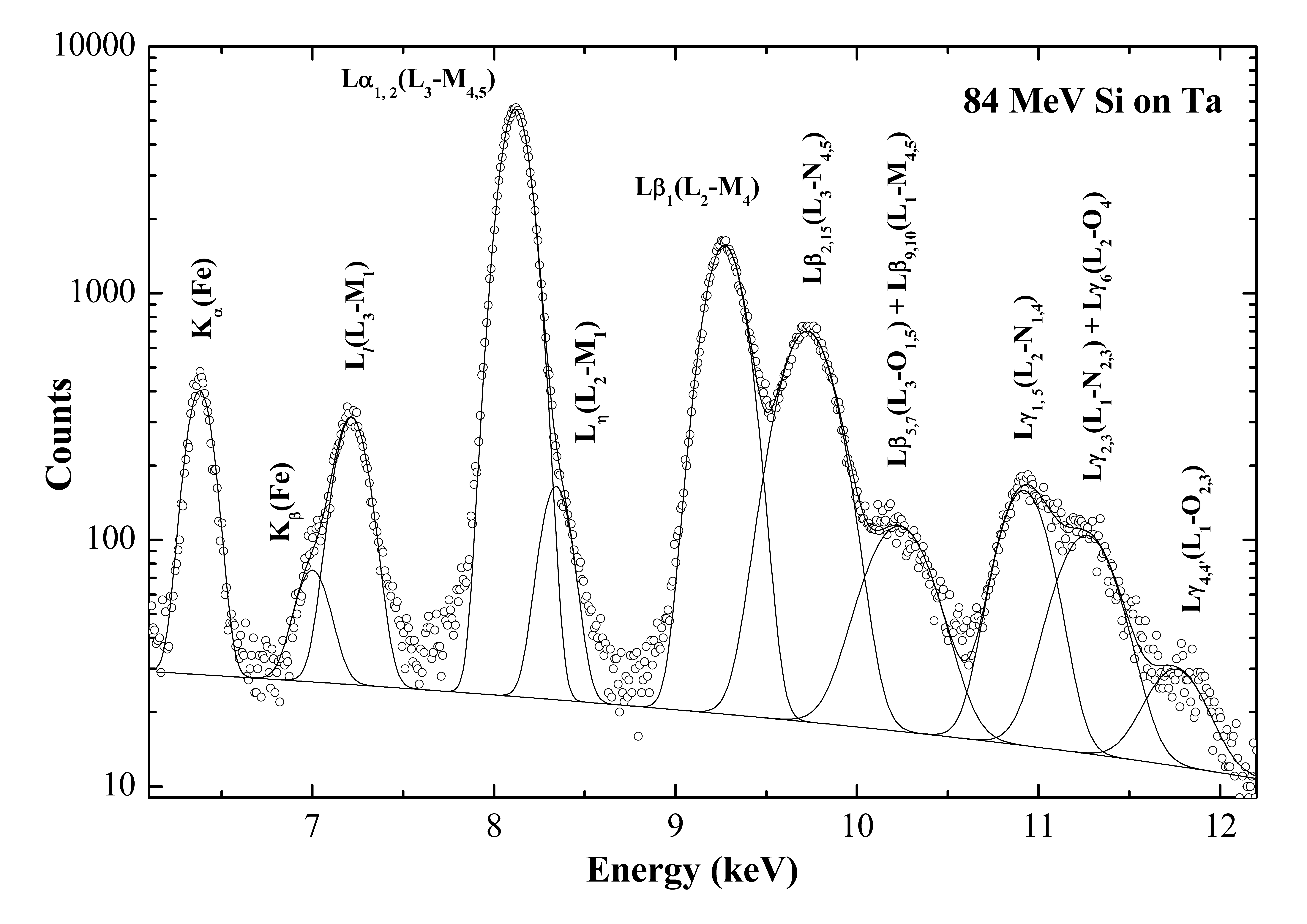}
\caption{\label{fig:1} L x-ray spectra of $_{73}{Ta}$ bombarded with the 84 MeV $^{28}{Si}$ ions. Deconvoluted x-ray lines due to different transitions are shown along with the background due to Compton scattering.}
\end{figure*}
\section{Measurement uncertainty evaluation method and results}
\indent A representative spectrum of emitted L x-rays induced by heavy particles is shown in Fig.~\ref{fig:1}. Thus, we determine the experimentally x-ray production cross-section for a particular x-ray peak ($\sigma_x^i$) using a relation
\begin{eqnarray}
\sigma_x^i= \frac{Y_x^i A Sin\theta}{N_A \epsilon n_p t \beta}\label{eq:2}
\end{eqnarray}
\noindent where $Y_x^i$ is the intensity of the $i$th x-ray peak, $A$ is the atomic weight of the target, $\theta$ is the angle between the incident ion beam and the target foil normal, $N_A$ is the Avogadro number, $n_p$ is the number of incident projectiles, $\epsilon$ is the effective efficiency of the x-ray detector, $t$ is the target thickness in $\mu$g/cm$^2$ and $\beta$ is a correction factor for the absorption of the emitted x-rays inside the target. Measurement uncertainty on the $\sigma_x^i$ is introduced due to uncertainties in different parameters used in Eqs.~(\ref{eq:2}). Let us consider an experiment \cite{OSWAL2020108809} to have ideas on the uncertainties involved, which are as follows: the photo peak area, $Y_x^i$, evaluation ($\leq1\% $ for the L$_\alpha$ x-ray peak and $\approx$ 3\% for the other peaks), the ion beam current ($\approx 5\% $), ${Sin\theta}$ ($\approx  1\% $) and the target thickness ($\approx$ 3\%). The uncertainty in the detector efficiency values, $\epsilon$, is $\approx$ 8\% in the energy region of current interest. The relative combined standard uncertainty equation for the x-ray production cross-section as given in Eqs.~(\ref{eq:2}) can be written including the relative combined variance and covariance as follows
\begin{equation}
\begin{aligned}
{(\frac{\delta\sigma^i_{x}}{\sigma^i_{x}})}^2= {(\frac{{\delta Y{_x^i}}}{Y{_x^i}})}^2+{(\frac{{\delta\epsilon}}{\epsilon})}^2+{(\frac{{\delta{n_p}}}{n_p})}^2+ {(\frac{{\delta{t}}}{t})}^2+{(\frac{{\delta{Sin\theta}}}{Sin\theta})}^2+{(\frac{{\delta{\beta}}}{\beta})}^2\\
+ 2~{\delta t}~{\delta \beta}~\frac{\delta \sigma^i_x}{\delta t}~\frac{\delta \sigma^i_x}{\delta\beta} \frac{1}{{\sigma^i_x}^2} 
- 2~{\delta \epsilon}~{\delta Y{_x^i}}~\frac{\delta \sigma^i_x}{\delta\epsilon}~\frac{\delta \sigma^i_X}{\delta Y{_x^i}} \frac{1}{{\sigma^i_x}^2}
\label{eq:3}
\end{aligned}
\end{equation}

\begin{table}
\caption{\label{tab:table2}
X-ray production cross-sections with uncertainty for the elements of our interest (uncertainties are  obtained using the uncertainties on various parameters mentioned in \cite{OSWAL2020108809} and given in terms of percentage).}
\centering
\begin{tabular}{l l l l}
\hline
 \textbf{El.} & \textbf{$\sigma_{L_{\gamma_{2+3}}}$} & \textbf{$\sigma_{L_{\gamma_{1+5}}}$} & \textbf{$\sigma_{L_{\alpha_{1+2}}}$}\\
\hline
$_{73}{Ta}$ & 193 $\pm$ 12 & 290 $\pm$ 12 & 6614 $\pm$ 11\\
$_{78}{Pt}$	& 154 $\pm$ 12& 231 $\pm$ 12& 5542 $\pm$ 11 \\
$_{90}{Th}$ & 42 $\pm$ 12& 63 $\pm$ 12 & 1537 $\pm$ 11\\
$_{92}{U}$	& 28 $\pm$ 12& 42 $\pm$ 12& 1098 $\pm$ 11\\
\hline
\end{tabular}
\end{table}

{\noindent Last two terms in uncertainty arise due to the contributions coming from the covariance relationship of the detector efficiency ($\epsilon$) with the $i^{th}$ intensity peak ($Y_i^x$) and the absorption correction factor of the emitted x-rays inside the target ($\beta$) with the target thickness (t). 
Simplifying these two terms by the equation (\ref{eq:2}) we can write the equation (\ref{eq:3}) as under
\begin{equation}
\begin{aligned}
{(\frac{\delta\sigma^i_{x}}{\sigma^i_{x}})}^2= {(\frac{{\delta Y{_x^i}}}{Y{_x^i}})}^2+{(\frac{{\delta\epsilon}}{\epsilon})}^2+{(\frac{{\delta{n_p}}}{n_p})}^2+ {(\frac{{\delta{t}}}{t})}^2+{(\frac{{\delta{Sin\theta}}}{Sin\theta})}^2+{(\frac{{\delta{\beta}}}{\beta})}^2\\+2~\frac{\delta \epsilon}{\epsilon}~\frac{{\delta Y{_x^i}}}{Y{_x^i}}+2~\frac{\delta t}{t}~\frac{\delta \beta}{\beta}
\label{eq:4}
\end{aligned}
\end{equation}
}
The absorption correction factor $\beta$ is written as

\begin{eqnarray}
\beta= \frac{1-\exp^{-\mu t}}{\mu t}
\label{eq:5}
\end{eqnarray}
and  
\begin{eqnarray}
\frac{\delta \beta}{\beta}\approx -\mu t~ (\frac{\delta t}{t}+\frac{\delta\mu}{\mu}).
\label{eq:6}
\end{eqnarray}
Where, $\mu$ is for attenuation co-efficient inside the target. The unit of $\mu$ is $cm^2/g$ and that of $t$ is $\mu g/cm^2$; the value $\mu t$ is very small. For instance, $\mu$ of carbon  for Fe $K_\alpha$ energy (6.4 keV) is 8.73 $cm^2/g$, $\frac{\delta\mu}{\mu}=0.5\%$ \cite{creagh1990problems} and if the target thickness is taken as 50 $\mu$g/cm$^2$ and $\frac{\delta t}{t}=3\%$, we get $\frac{\delta\beta}{\beta}=0.0013\%$. Thus, $\frac{\delta \beta}{\beta}$ is negligibly small and the terms containing $\delta\beta$ can be equated to zero. With this circumstances, the equation (\ref{eq:4}) can be written as  
\begin{equation}
\begin{aligned}
{(\frac{\delta\sigma^i_{x}}{\sigma^i_{x}})}^2= {(\frac{{\delta Y{_x^i}}}{Y{_x^i}})}^2+{(\frac{{\delta\epsilon}}{\epsilon})}^2+{(\frac{{\delta{n_p}}}{n_p})}^2+ {(\frac{{\delta{t}}}{t})}^2+{(\frac{{\delta{Sin\theta}}}{Sin\theta})}^2+2~\frac{\delta \epsilon}{\epsilon}~\frac{{\delta Y{_x^i}}}{Y{_x^i}}
\label{eq:7}
\end{aligned}
\end{equation}
\\
\indent Here we show that the $\epsilon$ does not have any relation with any quantities present in equation (\ref{eq:2}). The detector efficiency ($\epsilon$) is written according to a semi empirical model \cite{mohanty2008comparison} as
\begin{equation}
\begin{aligned}
\epsilon=\frac{\Omega}{4\pi}exp(-\sum_1^n \mu_id_i)f_E(1-exp(-\mu_iD))
\label{eq:8}
\end{aligned}
\end{equation}
where $\Omega$ is the fractional solid angle subtended by the detector crystal at the source and i denotes the medium between the source and the beryllium window, the window itself, a possible ice layer, the gold electrode, the frontal crystal dead layer etc. The $d_i$ are the thickness of these absorbers and $\mu_i$ are their linear absorption coefficients taken from XCOM database \cite{berger2010xcom}. The term $f_E$ represents the escape peak correction factor. The $\mu_C$ is the photoelectric absorption coefficient of the detector crystal of thickness $D$. There is no quantity present in RHS of equation (\ref{eq:2}) is present in the above equation. Hence, $\epsilon$ does not have obvious covariance with another quantity. However, the magnitude of $Y^x_i$ is strongly affected by the value of $\epsilon$ and $\epsilon=f(E)$, $E$=energy of x-ray, hence, the $\epsilon$ is having relation with $Y^x_i$. Whereas $Y^x_i$  is directly proportional to the number of projectiles ($N_p$) as well as the target thickness ($t$). Hence, $\epsilon$ does have covariance with the $Y^x_i$  and no covariance with either $N_p$ and $t$.\\
\indent The measured L x-ray production cross-sections are given in Table~\ref{tab:table2}. To calculate the ionization cross-section from the x-ray production cross-section, we need appropriate relation between these two. It is a simple relation for the K shell case involving only fluorescence yield: $\sigma_K = \sigma_x\omega_K $, where $\sigma_K$ is the K shell ionization cross-section, $\sigma_x$ is the K x-ray production cross-section, $\omega_K$ is the K x-ray fluorescence yield \cite{tiracsouglu2008determination}. However, the ionization cross-sections of different L subshells \cite{singh2000subshell,rahangdale2016spectroscopic} are related to the x-ray production cross-sections through the following  relations involving several atomic parameters like the fraction of radiative transition rates, fluorescence yields corresponding to $L_i$ subshells ($\omega_i$) and Coster-Kronig transition probability ($f_{ij}$) between two subshells $i$ and $j$ is written as follows

\begin{eqnarray}
\sigma_{L_1}= \frac{\sigma_{L_{\gamma_{2+3}}}}{{\omega_1}{S_{\gamma_{2+3,1}}}}.\label{eq:9}
\end{eqnarray}

\noindent \textcolor{red}{where} $\sigma_{L_1}$ is the L$_1$ subshell ionization cross-section, ${S_{\gamma_{2+3,1}}}$ is the fraction of radiative transition rates for ${L_{\gamma_{2+3}}}$ and ${\omega_1}$ is  fluorescence yield  of all the transitions terminating at L$_1$. Similarly, the L$_2$ subshell ionization cross-section $\sigma_{L_2}$ is given as

\begin{eqnarray}
\sigma_{L_2}= \frac{\sigma_{L_{\gamma_{1+5}}}}{{\omega_2}{S_{\gamma_{1+5,2}}}}-\sigma_{L_1}f_{12}.\label{eq:10}
\end{eqnarray}

\noindent Where ${S_{\gamma_{1+5,1}}}$ is the fraction of radiative transition rates for ${L_{\gamma_{1+5}}}$, ${\omega_2}$ is  fluorescence yield  of all the transitions terminating at L$_2$ and $f_{12}$ is the Coster-Kronig transition because of vacancy transfer from L$_1$ to L$_2$. Lastly, the L$_3$ subshell ionization cross-section $\sigma_{L_3}$ is given as
\begin{eqnarray}
\sigma_{L_3}= \frac{\sigma_{L_{\alpha_{1+2}}}}{{\omega_3}{S_{\alpha_{1+2,3}}}}-\sigma_{L_1}(f_{12}f_{23}+f_{13})-\sigma_{L_2}f_{23}.\label{eq:11}
\end{eqnarray}
\noindent Where ${S_{\alpha_{1+2,1}}}$ is the fraction of radiative transition rates, ${L_{\alpha_{1+2}}}$, ${\omega_3}$ is  fluorescence yield  of all the transitions terminating at L$_3$, $f_{13}$ is the Coster-Kronig transition because of vacancy transfer from L$_1$ to L$_3$, and $f_{23}$ is the Coster-Kronig transition because of vacancy transfer from L$_2$ to L$_3$.

\indent We can notice that obtaining ${\sigma_{L_1}}$ from Eqs.~(\ref{eq:9}) is simpler than finding ${\sigma_{L_{2,3}}}$ from Eqs.~(\ref{eq:10}) and Eqs.~(\ref{eq:11}), respectively, as Eqs.~(\ref{eq:8}) does not contain any Coster-Kronig transition probability. Whereas Eqs.~(\ref{eq:10}) and Eqs.~(\ref{eq:11}) contain one $f_{ij}$ and three $f_{ij}$s, respectively. The value of ${\omega}$’s and f’s are taken from Campbell et al. \cite{campbell2003fluorescence,campbell2009fluorescence} and the value of emission rates are taken from \citet{campbell1989interpolated} for obtaining the fraction of radiative transition rates. The emission rates of $L_1$, $L_2$ and $L_3$ subshells for the elements of our interest are given below in Table~\ref{tab:table3}, where, $L_1M_3$ etc., denote the origin of x-ray transition. For example, $L_1M_3$ is due to an electron jumping from $M_3$ to vacant $L_1$ subshell. $L_3O_{45}$ denotes a pair of emissions from $O_{45}$ to vacant $L_1$ subshell having line energies very close. For all the emission rate data, the uncertainty reported \cite{campbell1989interpolated} is 0.2\%. Using these emission rates, we can calculate the fraction of radiative transition rates when two or more lines are measured as a single line due to limited resolving power of detector used. Since $L_{\gamma_{2+3}}$ line is due to $L_{\gamma_{2}}$ ($L_1N_2$) and $L_{\gamma_{3}}$ ($L_1N_3$) transitions, so the fraction of radiative transition rates for $L_{\gamma_{2+3}}$ line is written as

\begin{eqnarray}
S_{\gamma_{2+3,1}}=\frac{L_1N_2+L_1N_3}{\sum_{i}{L_1X_i}}=\frac{L_1N_2}{\sum_{i}{L_1X_i}}+\frac{L_1N_3}{\sum_{i}{L_1X_i}}
\label{eq:12}
\end{eqnarray}

\noindent Now, uncertainty equation for the ${S_{\gamma_{2+3,1}}}$ can be written as

\begin{eqnarray}
{(\frac{{\delta{S_{\gamma_{2+3,1}}}}}{S_{\gamma_{2+3,1}}})}^2= {(\frac{{\delta{L_1N_2}}}{L_1N_2})}^2+{(\frac{{\delta{L_1N_3}}}{L_1N_3})}^2+2
{(\frac{{\sum_{i}\delta L_1X_i}}{{\sum_{i}L_{1}{X_i}}})}^2 \label{eq:13}
\end{eqnarray}
\noindent Similarly, we can calculate $S_{\gamma_{1+5,2}}$ and $S_{\alpha_{1+2,3}}$. Since $L_{\gamma{1+5}}$ line is due to $L_{\gamma{1}}$ $(L_2N_4)$ and $L_{\gamma{5}}$ $(L_2N_1)$ transitions so the fraction of radiative transition rates for for $L_{\gamma_{1+5}}$ line is

\begin{eqnarray}
S_{\gamma_{1+5,1}}
= \frac{L_2N_4+L_2N_1}{\sum_{i}{L_2Y_i}}\label{eq:14A}
=\frac{L_2N_4}{\sum_{i}{L_2Y_i}}+\frac{L_2N_1}{\sum_{i}{L_2Y_i}}.\label{eq:14B}
\end{eqnarray}
\noindent $L_{\alpha_{1+2}}$ line is due to $L_{\alpha_{1}}$($L_3M_4$) and $L_{\alpha_{2}}$($L_3M_5$) transitions so the fraction of radiative transition rates for $L_{\alpha_{1+2}}$ line is:

\begin{eqnarray}
S_{\alpha_{1+2,3}} = \frac{L_3M_4+L_3M_5}{\sum_{i}{L_3Z_i}} \label{eq:15A}
=\frac{L_3M_4}{\sum_{i}{L_3Z_i}}+\frac{L_3M_5}{\sum_{i}{L_3Z_i}}.
\label{eq:15B}
\end{eqnarray}
\noindent All calculated fraction of radiative transition rates for elements of our interest are given in Table~\ref{tab:table4}. Uncertainty equation for $S_{\gamma_{1+5,1}}$ and $S_{\alpha_{1+2,3}}$ can also be written in terms of Eqs.~(\ref{eq:11}). Emission rates and corresponding uncertainties are known \cite{campbell1989interpolated}, so we can calculate uncertainty in every fraction of radiative transition rates as shown in Table~\ref{tab:table4} for the elements of our interest. 

\begin{table}
\caption{\label{tab:table3}
$L_1$, $L_2$ and $L_3$ emission rates for electric dipole transitions \cite{campbell1989interpolated} are listed with multiplication factor $10^{-2}$. All rates are primarily in $eV/\hbar$ ($=1.519\times$ ${10}^{15}$ $S^{-1}$), hence rates in the table are in unit of $1.519\times$ ${10}^{13}$ $S^{-1}$.}
\begin{tabular}{l l l l l l l l l l l}
\hline
 \textbf{El.} & \textbf{$L_1L_3$} & \textbf{$L_1M_2$} & \textbf{$L_1M_3$} & \textbf{$L_1N_2$} & \textbf{$L_1N_3$} & \textbf{$L_1O_{23}$} & \textbf{$L_1P_{23}$}& & \\ 
\hline
$_{73}{Ta}$ &	1.446 &	24.65 &	31.54 &	6.176 &	8.386 &	2.189\\
$_{78}{Pt}$	& 2.636 &	34.94 &	41.27 & 9.023 &	11.54 &	3.561\\
$_{90}{Th}$ &	10.82 &	75.68 &	69.68 &	20.75 &	21.87 &	l0.l0& 2.004 \\
$_{92}{U}$	& 13.62 & 	85.59 &	74.82 &	23.63 &	23.93 &	ll.56 & 2.146\\
\hline
\textbf{El.} & \textbf{$L_2M_1$} & \textbf{$L_2M_4$} & \textbf{$L_2N_1$} &\textbf{$L_2N_4$}&\textbf{$L_2O_1$} &\textbf{$L_2O_4$} & \textbf{$L_2P_1$}\\
\hline
$_{73}{Ta}$ &2.928	& 107.4&	0.7266&	21.02&	0.1256&	0.4713&	0.01584\\
$_{78}{Pt}$& 4.061&	148.0&	l.038&	30.63 &	0.2039&	2.772&	0.01050\\
$_{90}{Th}$ & 8.374&	295.1&	2.283&	68.56 &	0.5646 &	13.35 &	0.1240\\
$_{92}{U}$& 9.375 &	328.0&	2.580 &	77.63 &	0.6538 &	15.59 &	0.1397\\
\hline
\textbf{El.} &\textbf{$L_3M_1$} & \textbf{$L_3M_4$} & \textbf{$L_3M_5$} & \textbf{$L_3N_1$} &\textbf{$L_3N_4$} & \textbf{$L_3N_5$} & \textbf{$L_3O_1$} & \textbf{$L_3O_{45}$} & \textbf{$L_3P_1$}\\
\hline
$_{73}{Ta}$ & 4.378 &	9.655 &	85.06 &	1.054 &	1.772 &	15.88 &	0.l809 &	0.3825 & 0.0227\\
$_{78}{Pt}$ & 6.377 &	13.11&	115.3 &	1.569 &	2.511 &	22.61 &	0.3051 &	2.176 &	0.01513\\
$_{90}{Th}$ & 14.67 &	25.02 &	219.5 &	3.773 &	5.l92 &	47.4O &	0.9264 &	9.983 &	0.2009\\
$_{92}{U}$& 16.72 &	27.58 &	241.8 &	4.314 &	5.796 &	53.13 &	1.076 &	11.50 &	0.2299\\
\hline
\end{tabular}
\end{table}

\begin{table}
\caption{\label{tab:table4} The fraction of radiative transition rates for the elements of our interest (uncertainties are given in percentage).}
\centering
\begin{tabular}{l l l l}
\hline
\textbf{El.}&  \textbf{$S_{\gamma_{2+3,1}}$}  & \textbf{$S_{\gamma_{1+5,2}}$} &\textbf{$S_{\alpha_{1+2,3}}$} \\
\hline
$_{73}{Ta}$&	0.1957 $\pm$ 1&	0.1639 $\pm$ 1&	0.8000 $\pm$ 1\\
$_{78}{Pt}$&	0.1997 $\pm$ 1&	0.1696 $\pm$ 1&	0.7831 $\pm$ 1\\
$_{90}{Th}$&	0.2020 $\pm$ 1 &	0.1824 $\pm$ 1&	0.7485 $\pm$ 1\\
$_{92}{U}$& 0.2021 $\pm$ 1 &	0.1848 $\pm$ 1&	0.7437 $\pm$ 1\\
\hline
\end{tabular}
\end{table}

Now, from Eqs.~(\ref{eq:8}), the uncertainty equation for $\sigma_{L_1}$ can be written as:

\begin{eqnarray}
{(\frac{{\delta\sigma_{L_1}}}{\sigma_{L_1}})}^2= {(\frac{{\delta\sigma_{L_{\gamma_{2+3}}}}}{\sigma_{L_{\gamma_{2+3}}}})}^2+{(\frac{{\delta\omega_{1}}}{\omega_1})}^2+{(\frac{{S_{\gamma_{2+3,1}}}}{S_{\gamma_{2+3,1}}})}^2 \label{eq:16}
\end{eqnarray}

\noindent Value of $\sigma_{L_{\gamma_{2+3}}}$ (barns/atom), $\omega_1$ and $S_{\gamma_{2+3,1}}$ with uncertainties, required for calculation of uncertainty of $\sigma_{L_1}$, are given in Table~\ref{tab:table5}. Uncertainty of {$\sigma_{L_{\gamma_{2+3,1}}}$} is taken from Table~\ref{tab:table2}, that of $\omega_1$ is taken from \cite{campbell2003fluorescence,campbell2009fluorescence} and that of ${S_{\gamma_{2+3,1}}}$ is taken from Table~\ref{tab:table4}. Values of the $\sigma_{L_1}$ along with the uncertainties are given in Table~\ref{tab:table5}.

\begin{table}
\caption{\label{tab:table5} $\sigma_{L_1}$ (barns/atom) along with uncertainties for various target atoms by Si beams at 84 MeV. Corresponding values of $\sigma_{L_{\gamma_{2+3}}}$ (barns/atom) \cite{OSWAL2020108809}, conversion parameters ${\omega_1}$ \cite{campbell2003fluorescence, campbell2009fluorescence} and {$S_{\gamma_{2+3,1}}$} are also given (uncertainties given in percentage).}
\centering
\begin{tabular}{l l l l l}
\hline
\textbf{El.} & \textbf{$\sigma_{L_{\gamma_{2+3}}}$}&\textbf{${\omega_1}$}& \textbf{$S_{\gamma_{2+3,1}}$} & \textbf{$\sigma_{L_1}$}\\ \hline
$_{73}{Ta}$ & 193 $\pm$ 12 &	0.144 $\pm$ 15 & 0.1957 $\pm$ 1 & 6858 $\pm$ 19\\
$_{78}{Pt}$ &  154 $\pm$ 12 &	0.114 $\pm$ 30 &0.1997 $\pm$ 1 & 6756 $\pm$ 32\\
$_{90}{Th}$ & 42 $\pm$ 12 &	0.159 $\pm$ 35 & 0.2020 $\pm$ 1  & 1308 $\pm$ 37\\
$_{92}{U}$ & 28 $\pm$ 12 &	0.168 $\pm$ 35 & 0.2021 $\pm$ 1 & 819 $\pm$ 37 \\
\hline
\end{tabular}
\end{table}

Now, Eqs.~(\ref{eq:10}) can be written as 

\begin{eqnarray*}
\sigma_{L_2}= \sigma^A_{L_2}+\sigma^B_{L_2}
\end{eqnarray*}

\noindent where,

\begin{eqnarray*}
\sigma^A_{L_2}= \frac{\sigma_{L_{\gamma_{1+5}}}}{{\omega_2}{S_{\gamma_{1+5,2}}}},  \sigma^B_{L_2}=-\sigma_{L_1}f_{12}
\end{eqnarray*}

\noindent So, the uncertainty equation for ${\sigma_{L_2}}$ can be written as

\begin{eqnarray}
{(\frac{\delta\sigma_{L_2}}{\sigma_{L_2}})}^2 = {{(\frac{\sigma^A_{L_2}}{\sigma_{L_2}})}^2} {(\frac{\delta\sigma^A_{L_2}}{\sigma^A_{L_2}})}^2+{(\frac{\sigma^B_{L_2}}{\sigma_{L_2}}^2 )}{(\frac{\delta\sigma^B_{L_2}}{\sigma^B_{L_2}})}^2  \label{eq:17}    
\end{eqnarray}

\noindent where,

\begin{gather}
{(\frac{\delta\sigma^A_{L_2}}{\sigma^A_{L_2}})}^2 = {(\frac{\delta\sigma_{L_{\gamma_{1+5}}}}{\sigma_{L_{\gamma_{1+5}}}})}^2+{(\frac{\delta\omega_{2}}{\omega_{2}})}^2+{(\frac{\delta{S_{\gamma_{1+5,2}}}}{S_{\gamma_{1+5,2}}})}^2   \tag{17.1}\label{eq:17.1}
\end{gather}

\noindent and

\begin{gather}
{(\frac{\delta\sigma^B_{L_2}}{\sigma^B_{L_2}})}^2 = {(\frac{\delta\sigma_{L_1}}{\sigma_{L_1}})}^2 + {(\frac{\delta{f_{12}}}{{f_{12}}})}^2 \tag{17.2}\label{eq:17.2}
\end{gather}

\noindent Values of $\sigma_{L_{\gamma_{1+5}}}$ (barns/atom), $\omega_2$, $S_{\gamma_{1+5,2}}$ and $f_{12}$ with uncertainties, required for calculation of uncertainty for $\sigma_{L_2}$ are given in Table~\ref{tab:table6}. $\sigma_{L_1}$ along with its uncertainty is taken from Table~\ref{tab:table5}. 10\% uncertainty in production cross-section ({$\sigma_{L_{\gamma_{1+5,2}}}$}) is taken, which is due to the uncertainty in the photo-peak area evaluation ($\approx4\%$), the ion beam current ($\approx5\%$) and the target thickness ($\approx3\%$) \cite{OSWAL2020108809}. The uncertainty in the effective efficiency values is 5-8\% in the energy region of current interest \cite{OSWAL2020108809}, highest value 8\% is taken. Uncertainty for $\omega_2$ is 5\% for all element of our interest \cite{campbell2003fluorescence,campbell2009fluorescence}. For $f_{12}$, uncertainty for $_{73}{Ta}$ is 20\%, for $_{78}{Pt}$ is 30-40\%, for $_{90}{Th}$ is 50-100\% and for $_{92}{U}$ is 50-100\% \cite{campbell2003fluorescence,campbell2009fluorescence}. $\delta S_{{\gamma_{1+5,2}}}$ is due to uncertainty in the emission rate, which is only 0.2\% \cite{campbell1989interpolated}. Values of $\sigma_{L_2}$ with uncertainties is given in Table~\ref{tab:table6}.  

\begin{table}
\caption{\label{tab:table6} $\sigma_{L_2}$ (barns/atom) along with uncertainties for various target atoms by Si beams of 84 MeV. Corresponding values of $\sigma_{L_{\gamma_{1+5}}}$ (barns/atom) \cite{OSWAL2020108809}, conversion parameters ${\omega_2}$, $f_{12}$ \cite{campbell2003fluorescence, campbell2009fluorescence}) and {$S_{\gamma_{1+5, 2}}$} \cite{campbell1989interpolated} are also given (uncertainties given in percentage).}
\begin{tabular}{l l l l l l l l}
\hline
 \textbf{Elements} & \textbf{$\sigma_{L_{\gamma_{1+5}}}$}& \textbf {${\omega_2}$} &\textbf{$S_{\gamma_{1+5,2}}$}& \textbf{${f_{12}}$} & \textbf{$\sigma_{L_2}$}\\
\hline
$_{73}{Ta}$& 290 $\pm$ 12&	0.28 $\pm$ 5 &	0.1639 $\pm$ 1&0.118 $\pm$ 20 & 5510 $\pm$ 16\\
$_{78}{Pt}$& 231 $\pm$ 12&	0.344 $\pm$ 5&	0.1696 $\pm$ 1&0.075 $\pm$ 30-40 & 3448 $\pm$ 17\\
$_{90}{Th}$& 63 $\pm$ 12&	0.503 $\pm$ 5&	0.1824 $\pm$ 1&0.040$\pm$ 50-100 & 635 $\pm$ 15\\
$_{92}{U}$& 42 $\pm$ 12&	0.506 $\pm$ 5&	0.1848 $\pm$ 1&0.035 $\pm$ 50-100 & 417 $\pm$ 15\\
\hline
\end{tabular}
\end{table}

Similarly, the Eqs.~(\ref{eq:11}) can be written as

\begin{eqnarray*}
\sigma_{L_3}= \sigma^C_{L_3}+ \sigma^D_{L_3}+ \sigma^E_{L_3}+ \sigma^F_{L_3}
\end{eqnarray*}

\noindent So the uncertainty equation can be written as

\begin{equation}
\begin{aligned}
{(\frac{\delta\sigma_{L_3}}{\sigma_{L_3}})}^2 = {(\frac{\sigma^C_{L_3}}{\sigma_{L_3}})}^2 {(\frac{\delta\sigma^C_{L_3}}{\sigma^C_{L_3}})}^2+{(\frac{\sigma^D_{L_3}}{\sigma_{L_3}} )}^2{(\frac{\delta\sigma^D_{L_3}}{\sigma^D_{L_3}})}^2+{(\frac{\sigma^E_{L_3}}{\sigma_{L_3}})}^2 {(\frac{\delta\sigma^E_{L_3}}{\sigma^E_{L_3}})}^2+\\{(\frac{\sigma^F_{L_3}}{\sigma_{L_3}} )}^2{(\frac{\delta\sigma^F_{L_3}}{\sigma^F_{L_3}})}^2\label{eq:18}
\end{aligned}
\end{equation}

\noindent where,

\begin{eqnarray*}
{\sigma^C_{L_3} = \frac{\sigma_{L_{\alpha_{1+2}}}}{{\omega_3}{S_{\alpha_{1+2,3}}}}, 
\sigma^D_{L_3} = -\sigma{L_1}(f_{12}f_{23}),
\sigma^E_{L_3}=-\sigma{L_1}f_{13},
\sigma^F_{L_3}=-\sigma_{L_2}f_{23}}
\end{eqnarray*}

\noindent and the corresponding uncertainty equations as follows 

\begin{gather}
{(\frac{\delta\sigma^C_{L_3}}{\sigma^C_{L_3}})}^2={(\frac{\delta\sigma_{L_{\alpha_{1+2}}}}{\sigma_{L_{{\alpha_{1+2}}}}})}^2+{(\frac{\delta\omega_{3}}{\omega_{3}})}^2+{(\frac{\delta{S_{\alpha_{1+2,3}}}}{S_{\alpha_{1+2,3}}})}^2 \tag{18.1}\label{eq:18.1}
\end{gather}

\begin{gather}
{(\frac{\delta\sigma^D_{L_3}}{\sigma^D_{L_3}})}^2={(\frac{\delta\sigma_{L_1}}{\sigma_{L_1}})}^2 + {(\frac{\delta{f_{12}}}{{f_{12}}})}^2+{(\frac{\delta{f_{23}}}{{f_{23}}})}^2 \tag{18.2}\label{eq:18.2}
\end{gather}

\begin{gather}
{(\frac{\delta\sigma^E_{L_3}}{\sigma^E_{L_3}})}^2={(\frac{\delta\sigma_{L_1}}{\sigma_{L_1}})}^2 + {(\frac{\delta{f_{13}}}{{f_{13}}})}^2 \tag{18.3}\label{eq:18.3}
\end{gather}

\noindent and

\begin{gather}
{(\frac{\delta\sigma^F_{L_3}}{\sigma^F_{L_3}})}^2={(\frac{\delta\sigma_{L_2}}{\sigma_{L_2}})}^2 + {(\frac{\delta{f_{23}}}{{f_{23}}})}^2 \tag{18.4}\label{eq:18.4}
\end{gather}

\noindent Values of $\sigma_{L_{\alpha_{1+2}}}$, $\omega_3$, {$S_{\alpha_{1+2,3}}$}, $f_{23}$ and $f_{13}$, with uncertainties required to get the uncertainty in $\sigma_{L_3}$ are given in Table~\ref{tab:table7}. Values of $f_{12}$, $\sigma_{L_1}$ and $\sigma_{L_2}$ along with uncertainties are taken from Table~\ref{tab:table5} and Table~\ref{tab:table6}, respectively. The values of $\sigma_{L_3}$ along with the uncertainties are given in Table~\ref{tab:table7}. 

\begin{table}
\caption{\label{tab:table7} $\sigma_{L_3}$ (barns/atom) along with uncertainties for various target atoms by Si beams of 84 MeV. Corresponding values of $\sigma_{L_{\alpha_{1+2}}}$ (barns/atom) \cite{OSWAL2020108809}, conversion parameters (${\omega_3}$, $f_{23}$, $f_{13}$ \cite{campbell2003fluorescence, campbell2009fluorescence}) and $S_{\gamma_{1+5, 2}}$ \cite{campbell1989interpolated} are also given (uncertainties are given in percentage).}
\begin{tabular}{l l l l l l l l}
\hline
\textbf{El.} & \textbf{$\sigma_{L_{\alpha_{1+2}}}$} &\textbf{${\omega_3}$} & $S_{\alpha_{1+2, 3}}$& \textbf{${f_{23}}$}& \textbf{${f_{13}}$} & \textbf{$\sigma_{L_3}$}\\
\hline
$_{73}{Ta}$& 6614 $\pm$ 11&	0.251 $\pm$ 5 & 0.8000 $\pm$ 1&	0.134 $\pm$ 10&	0.328 $\pm$ 15 & 29844 $\pm$ 13\\	
$_{78}{Pt}$& 5542 $\pm$ 11&	0.303 $\pm$ 5 &0.7831 $\pm$ 1&	0.126 $\pm$ 10&	0.545 $\pm$ 20 & 19175 $\pm$ 16\\	
$_{90}{Th}$ & 1537 $\pm$ 11 & 0.424 $\pm$ 5 & 0.7485 $\pm$ 1& 0.103 $\pm$ 10&	0.62 $\pm$ 15& 3962 $\pm$ 17\\	
$_{92}{U}$ & 1099 $\pm$ 11&	0.444 $\pm$ 5&0.7437 $\pm$ 1& 0.14 $\pm$ 5&	0.62 $\pm$ 15& 2756 $\pm$ 16\\
\hline
\end{tabular}
\end{table}

\indent Lastly, total L-shell ionization ($\sigma_{L}$) and its uncertainty are given as under:

\begin{eqnarray}
\sigma_{L} = \sum_{i}\sigma_{L_i} = \sigma_{L_1} + \sigma_{L_2} + \sigma_{L_3}\label{eq:19}
\end{eqnarray}

\begin{eqnarray}
{(\frac{\delta\sigma_{L}}{\sigma_{L}})}^2={(\frac{\sigma_{L_1}}{\sigma_{L}})}^2{(\frac{\delta\sigma_{L_1}}{\sigma_{L_1}})}^2 + {(\frac{\sigma_{L_2}}{\sigma_{L}})}^2{(\frac{\delta\sigma_{L_2}}{\sigma_{L_2}})}^2 + {(\frac{\sigma_{L_3}}{\sigma_{L}})}^2{(\frac{\delta\sigma_{L_3}}{\sigma_{L_3}})}^2 \label{eq:20}
\end{eqnarray}

We have given $\sigma_{L}$ as a function of energies in Table~\ref{tab:table8}. Now, we know the measured $\sigma_{L_{1,2,3,L}}$ along with their correct uncertainties involved so that we can make use of these with right confidence.

\begin{landscape}
\begin{table}
\caption{\label{tab:table8} Values of total L shell ionization cross-section $\sigma_L$ (barns/atom) along with uncertainties of various target atoms by Si beams of different energies  (uncertainties are given in percentage).}
\begin{tabular}{l l l l l l l l}
\hline
\textbf{El.} & \textbf{84 MeV}& \textbf{90 MeV}& \textbf{98 MeV} &\textbf{107 MeV} &\textbf{118 MeV}& \textbf{128 MeV}&\textbf{140 MeV} \\
\hline
$_{73}{Ta}$& 42213 $\pm$ 10& 53086 $\pm$10	&64543 $\pm$ 10 	&76243 $\pm$ 10 	&167575 $\pm$ 10 & 191674 $\pm$ 10	&198142 $\pm$ 10\\
	
$_{78}{Pt}$& 29378 $\pm$ 13 &30303 $\pm$ 13	&38547 $\pm$ 13	&49786 $\pm$ 13	&95815 $\pm$ 13	&99213 $\pm$ 14&	107667 $\pm$15\\
	
$_{90}{Th}$ & 5905 $\pm$ 14  	& 7186 $\pm$ 14	& 9024 $\pm$ 14	& 13166 $\pm$ 14 &	21059 $\pm$ 14	&25499 $\pm$ 14 & 30782 $\pm$ 14\\
	
$_{92}{U}$ & 3992 $\pm$ 14	&4394 $\pm$ 14	&6173 $\pm$ 14	&7956 $\pm$ 14	&13377 $\pm$ 14 &15819 $\pm$ 14 	&20089 $\pm$ 14\\
\hline
\end{tabular}
\end{table}
\end{landscape}

\begin{figure}
\includegraphics[width=140mm,scale=01.0]{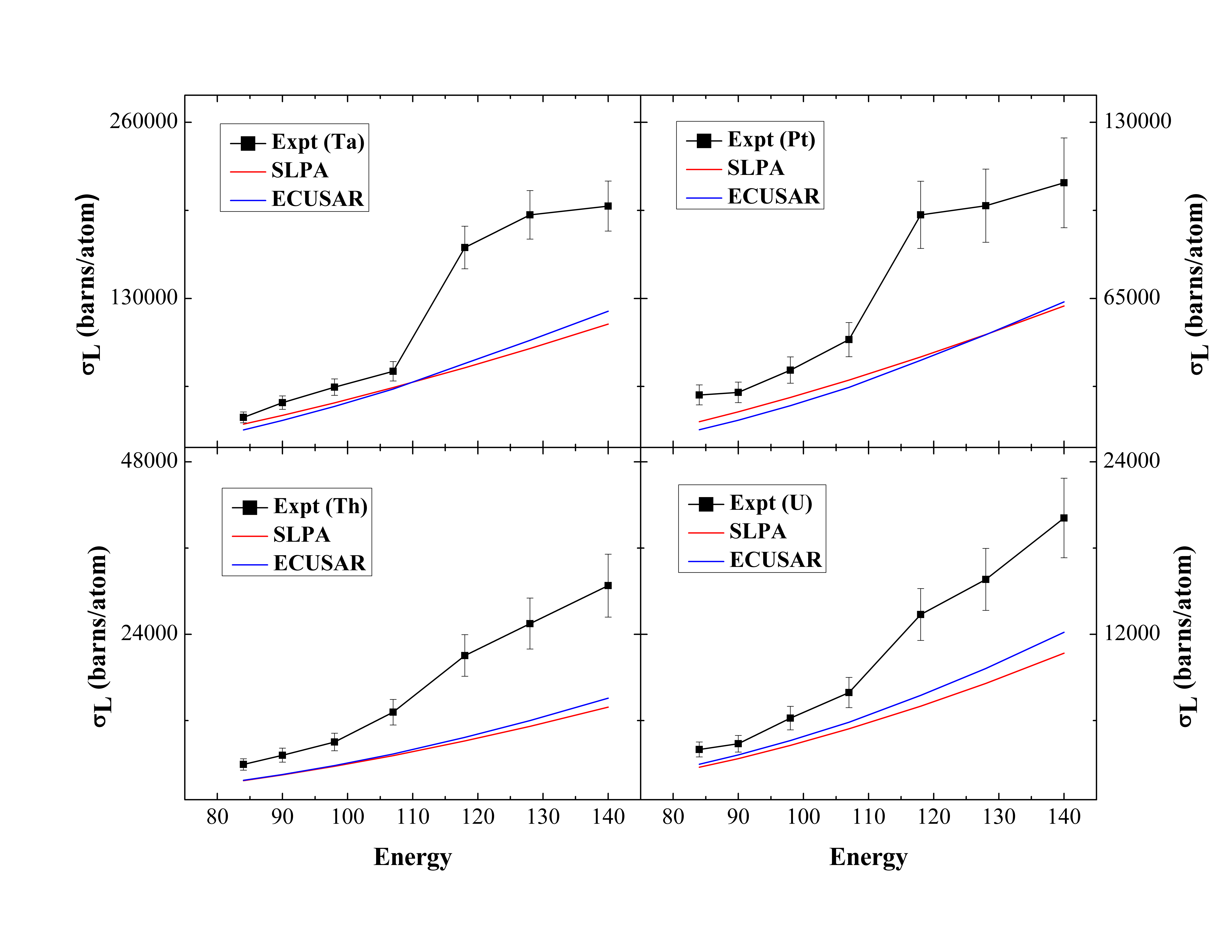}
\caption{\label{fig:2} Total ionization cross-section vs bombarding energy of Si ion on various target  Ta, Pt, Th and U. Measured data have been compared with ECUSAR \cite{lapicki2002status} and SLPA theory \cite{montanari2011collective}.}
\end{figure}

\section{Discussions on the salient results}
\indent The uncertainty in ${\sigma_{L_1}}$ as given in Table~\ref{tab:table5} for $_{73}{Ta}$ is about 19\%, where that of $_{78}{Pt}$ it is 32\%, and those of $_{90}{Th}$ and  $_{92}{U}$ it is about 37\%; even though the uncertainty in ${\sigma_{L_{\gamma_{2+3}}}}$ is about 12\%. \textcolor{red}{It is thus clear from the results that the uncertainty in ${\sigma_{L_1}}$ is primarily dictated by the large uncertainty in ${\omega_1}$ as given in Table~\ref{tab:table5}.} In the case of ${\sigma_{L_2}}$ the uncertainties are between $15-17\%$ for elements $_{73}{Ta}$, $_{78}{Pt}$, $_{90}{Th}$ and $_{92}{U}$ (Table~\ref{tab:table6}). It means the uncertainties are quite close to the uncertainty found in ${\sigma_{L_{\gamma_{1+5}}}}$ 12\%. It happens because of small uncertainty (5\%) in ${\omega_2}$. One important point is to note here that large uncertainty in ${f_{12}}$ does not affect the overall uncertainty in ${\sigma_{L_2}}$ because of weighted uncertainty propagation rule. We can notice that the value of ${f_{12}}$ is rather more important than the associated uncertainty because of the weighted factor changes with its contribution.\\
\indent The above trend seen with ${\sigma_{L_2}}$ plays a similar role in propagating the uncertainty in ${\sigma_{L_3}}$. Here also, the uncertainties are between $13-17\%$ for all four elements $_{73}{Ta}$, $_{78}{Pt}$, $_{90}{Th}$ and $_{92}{U}$. Since the first term containing fluorescence yield $\omega_3$ and the fraction of radiative transition rates $S_{\alpha_{1+2,3}}$ along with production cross-section ${\sigma_{L_{\alpha_{1+2}}}}$ in Eqs.~(\ref{eq:10}) gives major contribution, the uncertainty associated with this part is the main source of uncertainty. Further, uncertainty in $\omega_3$ is not much (5\%) and that in $S_{\alpha_{1+2,3}}$ is only 1\%, hence, the uncertainty in ${\sigma_{L_3}}$ is closed to that in ${\sigma_{L_{\alpha_{1+2}}}}$. Though the uncertainty in ${f_{23}}$ and ${f_{13}}$ are about 10 and 15\%, respectively, still their share in uncertainty propagation is small because their contributions in Eqs.~(\ref{eq:10}) (part 2 and 3), are small. Let us now list the uncertainty message on the L subshell ionization cross-sections reported earlier in Table \ref{tab:table9}.\\
\textcolor{red}{\indent Uncertainty in all L subshell ionization cross-sections cannot be equal, because the uncertainty in ${\sigma_{L_1}}$ is always larger due to larger uncertainty in ${\omega_1}$ than that in $\omega_{2}$ and $\omega_{3}$.  The uncertainty for $\sigma_{L_2}$ and $\sigma_{L_3}$ are nearly equal and highest uncertainty is with $\sigma_{L_1}$. These facts are well represented by the present work shown in Table \ref{tab:table9}. However, many cases shown in Table \ref{tab:table9} do not support these facts.  Hence, only some measurements treat the uncertainty associated with the L subshell ionization quite reasonably. This scenario is a deterrent for improving the theoretical understanding of the concerned complex mechanism. To give a correct treatment one has to follow elaborate error analysis as demonstrated in the present work. Having done this, one can confidently test the theories and look for any new physical processes if occurred in the experiment.} 
\begin{landscape}
\begin{table}
\tiny
\caption{\label{tab:table9} Comparison of uncertainties (\%) quoted for L shell ionization cross sections in the previous experiments and present study.}
\centering
\begin{tabular}{l l l l l l l}
\hline
\textbf{Ref.} & \makecell{Targets} & \makecell{\textbf{Uncertainty for}}& \makecell{\textbf{Projectile, its energy}} & \makecell{\textbf{Uncertainties (\%)}} & \makecell{\textbf{Remarks}}\\
\hline
\citet{palinkas1980subshell}& \makecell{Au, Pb, Bi} & \makecell{electron, 60-600 keV} &\makecell{$\sigma_{L_1}$, {$\sigma_{L_2}$}, {$\sigma_{L_3}$}} & \makecell{22, 15, 11} & \makecell{Fair treatment made}\\
\citet{PhysRevA.37.106} & \makecell{W} &\makecell{electron, 12-40 keV}&\makecell{$\sigma_3/\sigma_2$, $\sigma_2/\sigma_1$} &\makecell{10}& \makecell{{$\sigma_3/\sigma_2$} must have lower \\error than that in  {$\sigma_2/\sigma_1$}}\\ 
\citet{reusch1986method}& \makecell{$29\leq Z \leq79$} & \makecell{electron, 50-200 keV} & \makecell{$\sigma_{L_1}$, {$\sigma_{L_2}$}, {$\sigma_{L_3}$}, $\sigma_L$} & \makecell{7-15, 6-15, 6-15, 15} & \makecell{Error in $\sigma_{L_1}$ is underestimated}\\ 
\citet{bernstein1954shell}& \makecell{Ta, Au, Pb, U} & \makecell{H, 1.5-4.25 MeV} &\makecell{$\sigma_{L_x}$}& \makecell{20} & \makecell{No uncertainty message on L \\subshell ionization cross sections}\\ 
\citet{PhysRevA.11.607}& \makecell{Ta, Au, Bi} &\makecell{H, 1 to 5.5 and \\ He, 1 to 11 MeV} &\makecell{$\sigma_{L_1}$, {$\sigma_{L_2}$}, {$\sigma_{L_3}$} }& \makecell{22, 28, 21}& \makecell{Uncertainty for $L_2$ cannot \\ be higher than $L_1$ }\\ 

\citet{semaniak1994a}& \makecell{Between La to Au}& \makecell{ $^{14}N$, 1.75-22.4 MeV}&\makecell{$\sigma_{L_1}$, {$\sigma_{L_2}$}, {$\sigma_{L_3}$} }& \makecell{10-25\% for low energies \\and 10\% for higher energies}& \makecell{All subshells cannot \\have same uncertainty}\\
\citet{Semaniak1995b}& \makecell{$72\leq Z \leq 90$} &\makecell{C and N, 0.4 to 1.8 MeV/amu}&\makecell{$\sigma_{L_1}$, {$\sigma_{L_2}$}, {$\sigma_{L_3}$} }& \makecell{6-20, 5-12, 4-10} & \makecell{Achieving uncertainty as\\ low as 4\% for $\sigma_{L_3}$ is doubtful}\\
\citet{dhal1995subshell}& \makecell{Au, Bi}& \makecell{B, 4.8-8.8 MeV}& \makecell{$\sigma_{L_1}$, {$\sigma_{L_2}$}, {$\sigma_{L_3}$}} & \makecell{10\% for both elements.}& \makecell{All subshells cannot \\have same uncertainty}\\
\citet{banas2002role}& \makecell{Au} & \makecell{O of 0.4-2.2 MeV/amu}& \makecell{$\sigma_{L_1}$, {$\sigma_{L_2}$}, {$\sigma_{L_3}$}}&\makecell{15-30\%
for all subshell} & \makecell{Large error for $\sigma_{L_1}$ and small \\for $\sigma_{L_2}$ and $\sigma_{L_3}$is correct}\\
\citet{fijal2008subshell}&  \makecell{Au and Bi}& \makecell{S, 12.8-120 MeV} & \makecell{$\sigma_{L_1}$, {$\sigma_{L_2}$}, {$\sigma_{L_3}$}} &\makecell{10-30\% \\for all subshell}& \makecell{Large error for $\sigma_{L_1}$ and small \\for $\sigma_{L_2}$ and $\sigma_{L_3}$is correct}\\
\citet{OSWAL2020108809}&\makecell{Ta, Pt, Th, U}& \makecell{Si, 84-140 MeV}&\makecell{$\sigma_{L_1}$, {$\sigma_{L_2}$}, {$\sigma_{L_3}$}}& \makecell{15-20\% for Ta and\\ 30-35\% for others,\\ 12-15, 12-15}& \makecell{Good estimation}\\
Present work &\makecell{Ta, Pt, Th, U}& \makecell{Si, 84-140 MeV} & \makecell{$\sigma_{L}$} & \makecell{10\% for Ta, 13-15\% for \\Pt and 14\% for Th and U}& \makecell{correct treatment made}\\
Present work &\makecell{Ta}& \makecell{Si, 84 MeV} & \makecell{{$\sigma_{L1}$},{$\sigma_{L_2}$}, {$\sigma_{L_3}$}} & \makecell{19\%, 16\% and 13\% }& \makecell{Shown for a specific energy}\\
Present work &\makecell{Pt}& \makecell{Si, 84 MeV} & \makecell{{$\sigma_{L1}$},{$\sigma_{L_2}$}, {$\sigma_{L_3}$}} & \makecell{32\%, 17\% and 16\% }& \makecell{Shown for a specific energy}\\
Present work &\makecell{Th}& \makecell{Si, 84 MeV} & \makecell{{$\sigma_{L1}$},{$\sigma_{L_2}$}, {$\sigma_{L_3}$}} & \makecell{37\%, 15\% and 17\% }& \makecell{Shown for a specific energy}\\
Present work &\makecell{U}& \makecell{Si, 84 MeV} & \makecell{{$\sigma_{L1}$},{$\sigma_{L_2}$}, {$\sigma_{L_3}$}} & \makecell{37\%, 15\% and 16\% }& \makecell{Shown for a specific energy}\\
\hline
\end{tabular}
\end{table}
\end{landscape}
\indent The above methodology of error propagation can in principle be followed for future experiments. It can also be applied to earlier experiments if either data are still available for reanalysis or the production cross-section and its uncertainty are given in the concerned publications. However, the best available set of atomic parameters will be used to obtain L subshell ionization cross-sections from the L shell x-ray production cross-sections. \citet{barros2015ionization} have discussed this issue recently in great detail. The atomic parameters are the main source of uncertainty, especially the fluorescence yields. Till date, $\omega_2$ and $\omega_3$ are mostly known within 5\%, but $\omega_1$ is having large uncertainty in particular towards the heavier target side. This is the reason for the larger uncertainty in $\sigma_{L_1}$ in heavy atoms Th, U, etc. Though Coster-Kronig rates contain large uncertainty, but they do not affect much the L subshell cross-section because of their minor contribution (Eqs.~(\ref{eq:10})). Thanks to the theoretical advancements that the emission rates for the electric dipole transitions are known with almost no uncertainty (0.2\%). \\
\indent Plots of the total L shell ionization cross-section $\sigma_{L}$ along with uncertainty versus the projectile energy (Fig.~\ref{fig:2}) as given in Table~\ref{tab:table8} exhibit an unusual feature. The $\sigma_{L}$ behaves similarly, as the theoretical trend predicted by ECUSAR \cite{lapicki2002status} and SLPA \cite{montanari2011collective} up to a certain energy $\approx$ 100 MeV and takes a rise beyond this till it reaches a saturated value. Since we have dealt with the associated uncertainty in the measurement correctly, we can rely on the trend and infer on the concerned mechanism confidently. However, it is a nontrivial and requires a detail study that we keep for an interesting future investigation.\\
\section{Suggestion on making accurate L subshell cross section measurements}
\indent Of course on the experimental side, statistical uncertainty is an important issue for any measurements, in our case, it is L shell x-ray production cross-sections from which L subshell ionization cross-sections are derived. In this case, some other issues are more vital than the statistical uncertainty. For example, detector efficiency is the major source of uncertainty. Hence, final uncertainty on the L shell x-ray production cross-sections depend on the individual uncertainty on various measurements such as detector efficiency, counting statistics, number of incident ions and target thickness and target contamination. Let us discuss how the uncertainty of these experimental parameters can be kept minimum. We will make an attempt to find the best possible way of making the most accurate L shell x-ray production cross-section measurements below.\\
\indent First, we consider measurement of the detector efficiency of x-ray semiconductor detector used for x-ray production cross-sections. There are several efficiency measurements, for example, \citet{barfoot1984si,tribedi1992efficiency,mohanty2008comparison,kumar2017shell}. It is shown there that efficiency for a high purity germanium (HPGe) detector can be known within 5\% \cite{salgado2006determination}, for a Si(Li) detector within 6\% \cite{mohanty2008comparison}, and expected to get similar accuracy for a silicon drift detector (SDD) \cite{zhou2019multiple}. Hence, at this point, one can measure the detector efficiency within 5-6\%. Absolute efficiency of the detector depends not only on detector properties discussed but also on the correct position and solid angle called geometrical efficiency. To define it well, a collimator with a dimension smaller than the detector crystal size in front of the detector is normally used \cite{kumar2017shell}.\\
\indent Next, we discuss the measurement of the number of projectile-ions bombarded on the target. Since charge state of the heavy projectile-ions is changed to a good extent during the ion-solid collisions, charge collected in Faraday cage cannot give correct counting of incident particles. One must follow any one of the following techniques to have less than 1\% uncertainty: (i) elastically scattered projectiles from a heavy target like gold \cite{nandi2002lifetime}, keeping the whole scattering chamber electrically isolated for using it as a Faraday cup \cite{narvekar1992simple}, using two Faraday cups one in front of the target and another behind the target \cite{zhou2013k}. Target thickness, as well as target contamination, can be correctly measured with Rutherford back-scattering technique \cite{feldman1986fundamentals}.\\
\section{Crystal spectrometer vs semiconductor x-ray detector in measuring L subshell cross section}
\indent High-resolution studies using crystal spectrometers help us know preciously the characteristic line energies. However, such efforts do not improve the accuracy of the L subshell ionization because in high-resolution cases, we need only the emission rates, the fraction of radiative transition rates has no role over here. Since emission rates are very accurate (within 0.2 \%) and thus the fraction of radiative transition rates, as discussed above, does not introduce considerable uncertainty at all. Hence, experiments performed with crystal spectrometers, for example  \cite{reusch1986method,PhysRevA.37.106,llovet2014cross}, reported exactly similar uncertainty message as the studies using semiconductor x-ray detectors do. \\
\section{Conclusion}
\indent We always gain a good understanding of any physical process by comparing the experimental data with the theoretical values. The comparison is meaningful only if the uncertainty in experiment and theory is accounted to the best possible level. However, it is seldom seen that the theoretical works talk any thing about the uncertainty in the calculation. Often, we can find experimental studies quote these uncertainties, however, most of the time the uncertainties are not estimated correctly, especially for a complex phenomenon. We have taken a test case through L subshell ionization of atoms by particle or photon impact. Experimentally, measured x-ray production cross-sections are used to obtain the ionization cross-sections, which are theoretically calculated. Furthermore the uncertainty of the x-ray production cross-section is mainly governed by statistics (Type A uncertainty) and detector efficiency (Type B uncertainty) but the ionization cross-section involves many atomic parameters (systematic uncertainty) having a wide uncertainty spectrum. Consequently, determining the measurement uncertainty in the L subshell ionization cross-section is always quite complex. We have thoroughly studied this issue in a systematic manner, where the rule of the weighted propagation of the uncertainty is introduced so that on every occasion, percentage uncertainty can be used. We notice that $\sigma_{L_1}$ suffers from the maximum uncertainty because of a large uncertainty in the best fluorescence parameters (systematic uncertainty) available till date relevant to L$_1$ (2s$_{1/2}$) subshell compared to those associated with the other two subshells L$_2$ (2p$_{1/2}$) and L$_3$ (2p$_{3/2}$). \textcolor{red}{Hence, a comparison between the theory and experiment would give higher importance to $L_2$ and $L_3$ subshell ionisation cross-sections because of smaller measurement uncertainty than that in $L_1$.}\\
\indent Sometimes the measured data show a certain departure from the usual trend indicating certain unexplored mechanism. Nevertheless, a lack of confidence in associated uncertainties forbids one to unfold the mystery. At such juncture, log scale is used so that the data may look smoother than when it is on a linear scale, and thus any excessive scatter or unusual trend can be less noticeable. In contrast, accurate uncertainty analysis gives us full confidence about the data, and thus one will not hesitate to draw any meaningful inference if any departure is seen with the measured data. Furthermore, proper uncertainty analysis teaches us the best possible care that can be taken for the improved measurements. Thus, we have suggested a novel way by which the reliable and accurate L subshell ionization cross-section data can be obtained. Finally, we believe this work provides a general overview to every researcher that the uncertainty aspect must be taken into utmost care for right advancement of science.\\
\section{Credit authorship contribution statement}

Shashank Singh: Methodology, calculation, preparation of Tables and figures, Soumya Chatterjee: Methodology, calculation, preparation of Tables and figures, D. Mitra: Formal analysis, methodology, validation, and T. Nandi: Conceptualization, methodology, supervision, validation, visualization, writing - original draft, writing - review and editing.

\section{Acknowledgments}
{One of the authors (S.S.) gratefully acknowledges his supervisors K.P. Singh, Mumtaz Oswal and B. R. Behera for their support to work with any other experts freely for widening his exposures. We gratefully acknowledge Adedamola David Aladese for checking the manuscript thoroughly and giving invaluable comments.}


\bibliographystyle{apsrev4-1}
\bibliography{els-temlet.bbl}
\end{document}